# Computation-resource-efficient Task-oriented Communications

Jingwen Fu, *Student Member, IEEE,* Ming Xiao, *Senior Member, IEEE,* Chao Ren, *Member, IEEE,* Mikael Skoglund *Fellow, IEEE*

*Abstract*—The rapid development of deep-learning enabled task-oriented communications (TOC) significantly shifts the paradigm of wireless communications. However, the high computation demands, particularly in resource-constrained systems e.g., mobile phones and UAVs, make TOC challenging for many tasks. To address the problem, we propose a novel TOC method with two models: a static and a dynamic model. In the static model, we apply a neural network (NN) as a task-oriented encoder (TOE) when there is no computation budget constraint. The dynamic model is used when device computation resources are limited, and it uses dynamic NNs with multiple exits as the TOE. The dynamic model sorts input data by complexity with thresholds, allowing the efficient allocation of computation resources. Furthermore, we analyze the convergence of the proposed TOC methods and show that the model converges at rate $O\left(\frac{1}{\sqrt{T}}\right)$ with an epoch of length $T$. Experimental results demonstrate that the static model outperforms baseline models in terms of transmitted dimensions, floating-point operations (FLOPs), and accuracy simultaneously. The dynamic model can further improve accuracy and computational demand, providing an improved solution for resource-constrained systems.

## I. Introduction

The seminal work by Weaver and Shannon identified three levels of communication: the first level concerns the accurate transmission of symbols; the second level addresses the conveyance of semantic meaning within context; and the third level focuses on effectively executing the tasks [1]. Traditionally, communication systems have concentrated primarily on the first level, i.e., ensuring the reliability of symbol transmission. Each component, such as the transmitter, receiver, encoder, and decoder, within the communication system, is separately designed and optimized based on various assumptions and transmission goals [2].

In recent years, the rapid development of deep learning has significantly changed the methods of wireless communications. In [3], the encoder and decoder were parameterized by convolutional neural networks (CNNs) and were jointly

The work is supported in part by European Commission, Horizon Europe, MSCA Project, SCION (Secured and Intelligent Massive Machine-to-Machine Communication for 6G), and ASCENT (Autonomous Vehicular Edge Computing and Networking for Intelligent Transportation). and in part by Swedish Research Council Project entitled "Coding for Large-scale Distributed Machine Learning", number 2021-04772. The computations were enabled by resources provided by the National Academic Infrastructure for Supercomputing in Sweden (NAISS), partially funded by the Swedish Research Council through grant agreement no. 2022-06725.

Jingwen Fu, Ming Xiao, Chao Ren, and Mikael Skoglund are with the School of Electrical Engineering and Computer Science (EECS), KTH Royal Institute of Technology, 11428 Stockholm, Sweden. (Corresponding author: Ming Xiao.) Email: {jingwenf, mingx, chaor, skoglund}@kth.se.

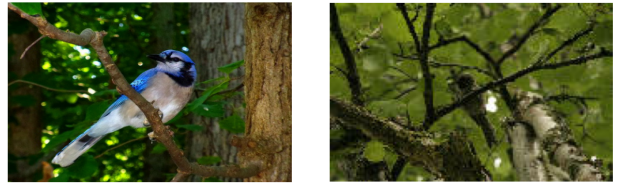

Fig. 1: Example of classifying a bird [7] with one simple image (left) and one complex image (right).

trained, representing a paradigm shift towards deep learning-enabled communication models. Furthermore, a deep learning-enabled semantic communications method, namely DeepSC, has explored the second level of communications [4].

Meanwhile, task-oriented communication (TOC) represents a higher level of communication, where the objective is to convey least task-related information while minimizing redundancy. For this purpose, the system needs to establish a shared knowledge base (KB) between the transmitter and receiver and apply joint training methods on both sides. This approach has several benefits: it improves communication efficiency by reducing unnecessary data transmission and increases privacy by transmitting representative features rather than raw data [5].

Despite the clear benefits of TOC, there are still severe challenges for TOC in practice:

(1) *Limited computation resources in devices.* In general, TOC systems need substantial computation resources to extract features and build the shared KB. For devices with powerful graphics processing units (GPUs), where there are abundant computation resources, large neural networks (NNs) can be used to improve performance. However, for devices such as uncrewed aerial vehicles (UAVs), mobile phones, or wearable monitoring sensors, the computation resources are in general limited. Thus, efficient methods exploiting limited computation resources are necessary [6]. Furthermore, it is expensive to train independent NNs for each device. Common NNs that can be used by different devices are preferable.

(2) *Diverse task complexity.* Tasks may differ in their computational complexities. For instance, in an image classification task, there are both simple and complex samples. The left image in Fig. 1 shows the front view of a bird with clear features, making it straightforward to classify. In contrast, the right image in Fig. 1 shows a bird hiding in the woods with a similar color to the tree trunk, which requires a more computationally intensive model to identify. Ideally, the TOC model should be able to automatically recognize the task complexity and allocate computation resources accordingly [8]. However, current TOC models have not considered varying task complexity yet, leading to inefficient computation resource allocation.

To address the aforementioned challenges, we propose computation-resource-efficient TOC models. The proposed TOC system can operate in two models: a static model for tasks without computation resource limitations, and a dynamic model designed to dynamically allocate computation resources within a specified device budget. The main contributions of our paper can be summarized as follows:

- We propose a novel TOC model with two operation models: a static model for large devices without computation limitations, and a dynamic model for small devices with different computation budgets. To reduce complexity and computation complexity simultaneously, we propose a simplified NN-based channel encoder and decoder for the wireless communication block to reduce the transmitted feature dimensions. The NNs of the channel encoder/decoder consist of a few simple layers with low computation demand, effectively reducing the floating point operations (FLOPs) in the wireless communication block. In the static model, we apply ResNet as the static task-oriented encoder (TOE), which can enhance feature extraction ability and model flexibility [9]. In experiments on the CIFAR-100 dataset, for static models using ResNet-20 as TOE, the proposed method reduces transmitted dimensions by up to $99.22\%$ and FLOPs by up to $89.26\%$ compared to baseline models. Meanwhile, the proposed static model also improves accuracy by up to $41.75\%$ in additive white Gaussian noise (AWGN) channels and $24.99\%$ in Rayleigh fading channels compared to baseline models.
- The proposed dynamic TOC model addresses the challenges of limited resources and diverse task complexity. It uses multi-exit dynamic NNs as the TOE. During the training phase of the dynamic TOC model, a confidence score is calculated for every input image at every exit. Then the encoder utilizes a validation dataset to determine the confidence score thresholds, and accordingly distributes images to different exits to meet the computational constraint. During the testing phase, the TOE outputs simple image features in early exits and complex image features in late exits based on confidence score, thereby allowing dynamic computation resource allocation based on task complexity. In the experiments on the CIFAR-100 dataset, the results show that the proposed dynamic model further enhances accuracy up to $58.75\%$ compared to baseline models with a lower computation load. We also show that the dynamic TOC model can operate at different computation budgets.
- We give the theoretical proof for the convergence of the NN-based TOC models using stochastic gradient descent (SGD) optimization. Experimental results further validate the convergence of our proposed TOC method in both static and dynamic models.

The rest of this paper is organized as follows. Section II reviews the related works. Section III presents the system model of the proposed TOC method. Section IV and Section V introduce the proposed static and dynamic TOC model respectively. Section VI gives the convergence analysis of the proposed method. Numerical results are shown in Section VII. Finally, Section VIII concludes the paper.

## II. LITERATURE REVIEW

In this section, we will provide a review of semantic communication and TOC. Then, we will briefly review dynamic NNs, which will be used in our system.

### A. Semantic and task-oriented communications

Semantic communication aims to convey intended meanings rather than exact data, thereby enhancing efficiency and relevance in information exchange [10]. Two major modeling paradigms have emerged: NN-based approaches and reinforcement learning (RL)-based strategies. DeepSC, a transformer-based model, exemplifies the NN-based method by recovering semantics within text [4], with MU-DeepSC extending this to multi-modal and multi-user settings [11]. RL-based methods, such as [12], optimize semantic similarity through interaction-driven learning. In parallel, TOC prioritizes the successful execution of intended tasks by transmitting only task-relevant information [5]. Within NN-based TOC methods, [13] explores the rate-distortion trade-off during wireless transmission using the variational information bottleneck (VIB) as a theoretical limit, while [14] extends this to improve robustness via the robust information bottleneck (RIB). TOC is also explored in complex settings, including multi-task [15] and multi-device [16] scenarios, as well as multi-modal learning with foundation models [17]. Beyond NN-based methods, RL is also applied to address scene classification challenges in TOC [18], and graph neural networks (GNNs) are utilized to enhance communication efficacy [19].

To improve the performance of semantic and TOC systems, optimized resource allocation has attracted lots of research interest. In reference [20], semantic spectral efficiency is introduced to optimize resource allocation for channel assignment. In [21], compression ratios and resource allocation are optimized jointly to maximize the probability of task success. Reference [22] proposes semantic-bit quantization for non-guided adaptive resource allocation. Reference [23] uses joint resource management over the pre-training and fine-tuning stages for edge learning. Reference [24] proposes a channel-adaptive model using attention mechanisms to enable adaptive channel state information compression. However, these studies primarily focus on the allocation of channel resources, while the allocation of computational resources has been largely overlooked. A separate line of work improves computational resource allocation by employing smaller NN models, such as in [6], which reduces model size to lower transmission overhead. In contrast, our approach first leverages a NN-based channel encoder/decoder architecture to reduce computational complexity and transmission load jointly, and further incorporates a dynamic model to enable adaptive allocation of computational resources based on task complexity.

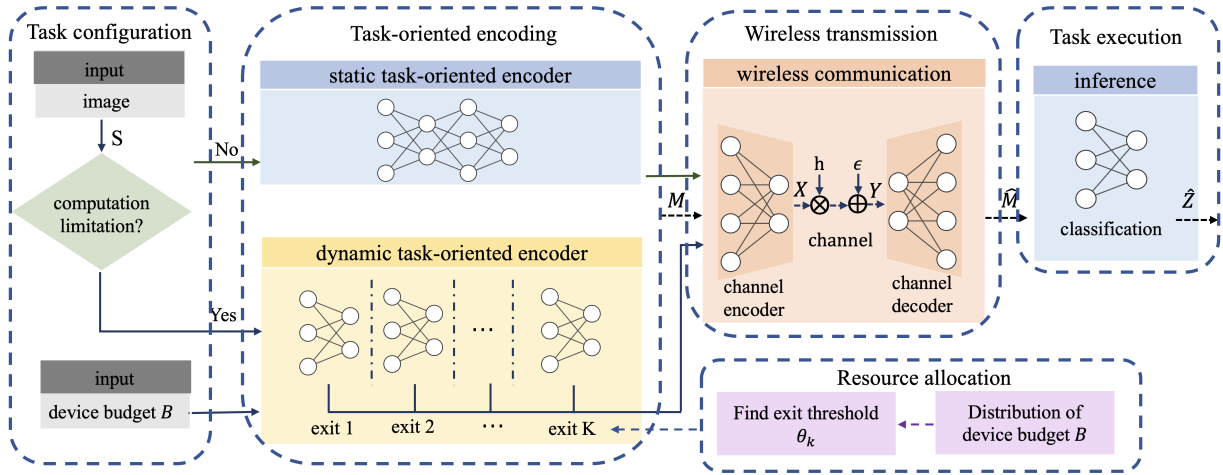

Fig. 2: System schematics of the proposed TOC models.

## B. Dynamic neural networks

Dynamic NNs have been extensively studied due to their efficiency, robust representational capabilities, and enhanced interpretability [25]. Roughly, dynamic NNs have dynamic architectures and can be categorized as dynamic depth, dynamic width, and dynamic routing methods. Despite advances in dynamic NNs, their applications in communication systems remain limited. In [13], dynamic width is utilized to adapt the size of the output features before transmission in channels. This approach differs from our work, where we implement dynamic depth tailored to the complexity of the input data. Reference [26] proposes an adaptive universal transformer network for semantic communications, which can be seen as utilizing dynamic depth in NNs. However, the performance gain remains marginal, as the performance of NNs is not simply proportional to NN depth.

## III. SYSTEM MODEL

The schematics of the proposed method can be found in Fig. 2, which includes task configuration, task-oriented encoding (TOE), wireless communications, and task execution blocks. Here we use an image classification task to illustrate the proposed TOC system, which operates in two models: static and dynamic. The task seeks to classify a batch of $N$ images with as high accuracy as possible. The system begins with the task configuration block, where the task setup is determined based on the computation capabilities of the transmitter devices. Specifically, when transmitters possess sufficient computational resources, such as GPU clusters, the static model is employed, leveraging large NNs to enhance classification performance. Conversely, if the transmitters have constrained computational capabilities, such as UAVs, mobile phones, or wearable sensors, the system adopts the dynamic model, wherein the model dynamically adjusts computational complexity according to the available resources.

Following the task configuration module, the TOE module processes data according to the selected model. The TOE receives a batch of $N$ images as input for the static encoder, and in the case of the dynamic encoder, it also receives an additional device computation budget $B$. This budget $B$, measured in FLOPs, quantifies the computational resources required, referring to the number of arithmetic operations performed on floating-point numbers. Detailed descriptions of the static and dynamic TOE implementations will be presented in Section IV and V, respectively.

The output of the TOE consists of compressed, task-relevant features, which are then passed into the wireless communication block for transmission. This transition constitutes a critical interface between on-device processing and wireless transmission. At this interface, the features produced by the TOE are first encoded by the channel encoder into a form suitable for transmission. These encoded signals are then transmitted over the wireless channel and subsequently decoded by the channel decoder to reconstruct the features for downstream tasks. The communication module employs an NN-based channel encoder and decoder to handle the transmission of features over noisy wireless channels. The channel is treated as NN with non-trainable parameters. Different from the VFE method, where an NN-based VIB structure is used to minimize the distance between the features transmitted before and after the channel [13], our system utilizes a simplified NN architecture as shown in Table I, consisting of two or three linear layers, each followed by non-linear activation layers for feature compression or decompression. Notably, our approach significantly reduces the transmitted feature dimensions (e.g., for CIFAR-100 dataset, our model compressed transmitted feature dimension down to 16, compared to 64 in VIB method [13] and 2048 in DeepJSCC method [3]). Additionally, compared to the complex VIB structure, which requires calculating KL divergence, the proposed NN-based channel encoder and decoder avoid such requirements, effectively reducing complexity and saving computation resources.

Finally, the decoded features produced by the wireless communication block are passed to the task execution block at the receiver side. In this stage, NNs perform inference based on the received features to complete the classification task at edge devices. For illustration, the whole TOC system can be

TABLE I
PROPOSED NN STRUCTURE FOR WIRELESS COMMUNICATION

| Layer Name | Type | Units | Activation |
|---|---|---|---|
| Channel Encoder | Linear | 128 | ReLU |
|  | Linear | 16 | None |
| Channel | AWGN/Rayleigh Fading | None | None |
| Channel Decoder | Linear (L1) | 16 | ReLU |
|  | Linear (L2) | 128 | ReLU |
|  | Linear (L3) | 16 | None |
|  | LayerNorm (L1+L3) | None | None |

expressed as the following process:

$$\mathcal{S} \xrightarrow{\beta} \mathcal{M} \xrightarrow{\eta} \mathcal{X} \xrightarrow{\text{channel}} \mathcal{Y} \xrightarrow{\kappa} \hat{\mathcal{M}} \xrightarrow{\gamma} \hat{\mathcal{Z}}, \quad (1)$$

where $\mathcal{S} = \{s_1, s_2, \ldots, s_j, \ldots, s_N\}$ represents a batch of $N$ source images, and $\mathcal{M} = \{m_1, m_2, \ldots, m_j, \ldots, m_N\}$ is the extracted task-related features, $\mathcal{X} = \{x_1, x_2, \ldots, x_j, \ldots, x_N\}$ denotes channel-encoded features, $\mathcal{Y} = \{y_1, y_2, \ldots, y_j, \ldots, y_N\}$ denotes received encoded features at the receiver, $\hat{\mathcal{M}} = \{\hat{m}_1, \hat{m}_2, \ldots, \hat{m}_j, \ldots, \hat{m}_N\}$ is the channel-decoded task-related features, and $\hat{\mathcal{Z}} = \{\hat{z}_1, \hat{z}_2, \ldots, \hat{z}_j, \ldots, \hat{z}_N\}$ is the final inference result.

The whole encoding process can be expressed as:

$$\mathcal{X} = \varepsilon_c(\varepsilon_t(\mathcal{S}; \beta); \eta), \quad (2)$$

where $\varepsilon_t(\cdot)$ denotes the TOE with trainable parameters $\beta$, and $\varepsilon_c(\cdot)$ denotes the channel encoder with trainable parameters $\eta$. The transmission process can be expressed as

$$\mathcal{Y} = h\mathcal{X} + \epsilon, \quad (3)$$

where $h$ and $\epsilon$ denote the channel gain and channel noise, respectively. On the receiver side, the decoding and task execution process can be expressed as

$$\hat{\mathcal{Z}} = \delta_t(\delta_c(\mathcal{Y}; \kappa); \gamma), \quad (4)$$

where $\delta_c(\cdot)$ denotes the channel decoder with trainable parameters $\kappa$ and $\delta_t(\cdot)$ denotes the task-oriented inference block with trainable parameters $\gamma$.

## IV. STATIC TASK-ORIENTED COMMUNICATION MODEL

In static TOC models, no computation limits are specified in the task-configuration phase. Therefore, there is no need to allocate computational resources for input images. In such model, a static TOE is applied for feature encoding. The static TOE takes source images $\mathcal{S}$ as inputs, and extracts task-related features $\mathcal{M}$ using NNs.

To improve performance and reduce complexity, we use ResNet of different sizes as the static TOE for feature extraction [9]. For the wireless transmission block, we use a simplified NN-based channel encoder and decoder for wireless communication to reduce the transmitted dimensions. As shown in Table I, the NNs of the channel encoder/decoder consist of a few simple layers with low computation complexity, effectively reducing the complexity of the wireless communication block. Furthermore, to improve performance, the static TOE NN is jointly trained with the NNs within the wireless transmission block and the inference block. For training, we use the cross-entropy loss for the classification task as follows,

$$\mathcal{L}_{static}(\mathcal{Z}, \hat{\mathcal{Z}}; \beta, \eta, \gamma, \kappa) = -\sum_{j=1}^{M} z_j \log(\hat{z}_j), \quad (5)$$

where $\mathcal{Z} = \{z_1, z_2, \ldots, z_j, \ldots, z_N\}$ is the true input image label, and $M$ is the number of classes in classification tasks.

## V. DYNAMIC TASK-ORIENTED COMMUNICATION MODEL

In dynamic TOC models, a transmitter device computation budget $B$ is specified in the task configuration phase. The dynamic TOE takes a batch of $N$ images $\mathcal{S}$ as inputs, then encodes them into task-related features $\mathcal{M}$. During this process, the encoder dynamically distributes the computation resources within the batch to satisfy the device budget limitation. Then our problem can be regarded as a budgeted batch classification problem, which will be discussed below.

The learning process of the dynamic TOC model has three stages, i.e., training, validation, and testing stages. First, the training stage produces a generalized dynamic NN suitable for devices with different budgets. In the subsequent validation phase, the model takes the device budget $B$ to determine the exit thresholds at each exit $k$ in the dynamic NNs. Finally, in the testing phase, the dynamic model assesses the confidence score of each input image while extracting image features, and outputs the feature if the score reaches a threshold. The details of these three stages will be presented as follows. Here we assume the training dataset $\mathcal{S}_{\text{train}}$, validating dataset $\mathcal{S}_{\text{val}}$, and testing dataset $\mathcal{S}_{\text{test}}$ to have similar distribution.

### A. Budgeted batch classification

For illustration, our dynamic TOC model can be expressed as a budgeted batch classification problem. A device computation budget $B$, is specified in the task configuration phase. The problem aims to assign more computation resources to complex input images and fewer resources to simpler input images while ensuring that the average computation cost remains within budget $B$ for a batch of $N$ images. In the testing stage, an image batch can be denoted as $\mathcal{S}_{\text{test}} = \{s_1, s_2, \cdots, s_j, \cdots, s_N\}$, where $s_j$ represents the $j$-th input image sample. The computation cost at exit $k$ is $C_k$, and the probability for outputting image feature $s_j$ at the $k$-th exit of the NN is $Pr_k(s_j)$. The average computation cost for the budgeted batch classification problem is constrained by

$$\sum_{k=1}^{K} \sum_{j=1}^{N} Pr_k(s_j) C_k \leq B, \quad s_j \in \mathcal{S}_{\text{test}}. \quad (6)$$

### B. The network architecture of dynamic TOE

To address the budgeted classification problem, we propose a multi-exit dynamic NN for the TOE. We note two main challenges in designing the network architecture: (1) Variation in information content across different depths. When using NNs to process images, normally, shallower layers process fine-scale and detailed features, whereas deeper layers deal with coarse-scale and generalized features [27]. Furthermore,

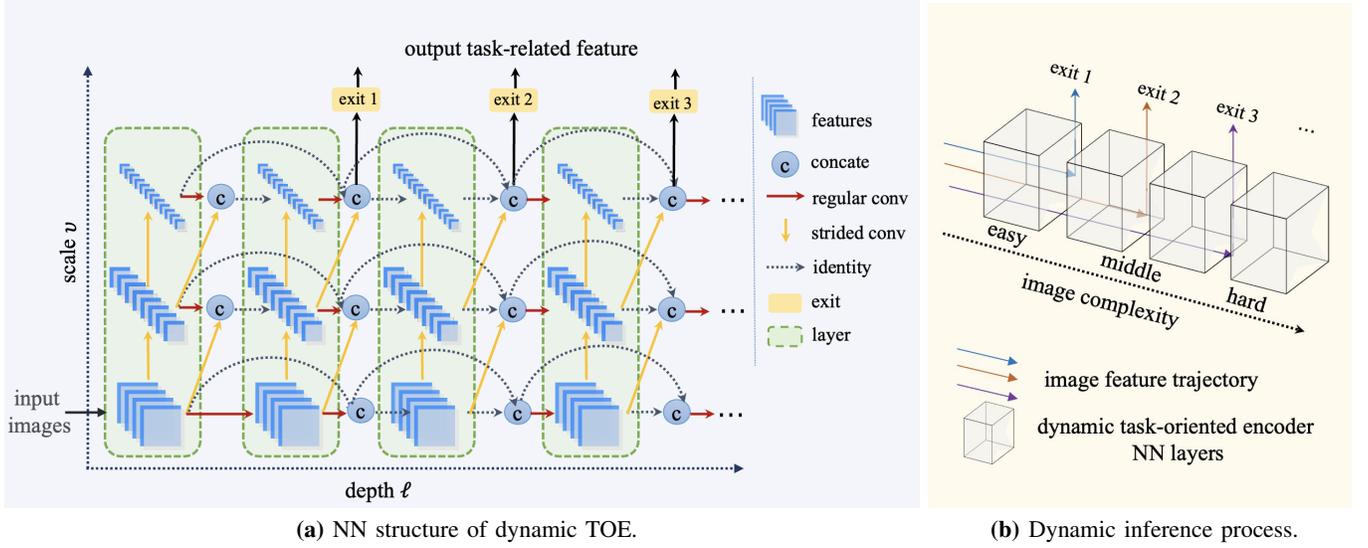

**(a)** NN structure of dynamic TOE.  **(b)** Dynamic inference process.

Fig. 3: (a) The NN structure of Dynamic TOE, which takes image samples as inputs, and extracts task-related features for channel encoding. The NN consists of multiple exit points for the dynamic allocation of resources. It has a multi-dimensional structure for the depth and scale of features respectively. (b) The inference process at the task execution side (the receiver). Simple image features are output from early exits, while complex image features are output from late exits.

the accuracy of classification tasks is notably related to the depth of the network. (2) Diverse feature scales at different exits. The scale of output features differs significantly across network depths [9]. This variation can lead to challenges for wireless transmission, as larger feature scales, especially in early exits, may increase processing time and cause delays, leading to higher transmission loads [13]. To solve the two challenges, we propose a multi-scale dynamic TOE with multiple exits, as shown in Fig. 3 (a). The main components of the TOE include multi-exits, dense connectivity, horizontal and vertical network structure, and confidence scoring. The multi-exit facilitates the encoder to terminate at different depths of NNs with different complexity. The dense connectivity effectively solves challenge (1), and the horizontal and vertical network structure addresses challenge (2). Finally, confidence scores denote the complexity of each input image and further facilitate the dynamic sorting and exiting process.

*1) Multi-exits:* An exit means a termination point in the dynamic NN structure. When reaching an exit, the extraction of image features stops, and the NN outputs features for further transmission. Confidence scores are evaluated at each exit point. The NNs of dynamic TOE have multiple exits, labeled from exit 1 to exit $K(K > 1)$, as depicted in Fig. 3 (a). The multi-exit NN allows early exiting for simple image features and late exiting for complex image features. In general, earlier exits require fewer computational resources, and later exits require more computational resources.

*2) Horizontal and vertical dimension of the NN structure:* The network structure of dynamic TOE has a horizontal dimension and a vertical dimension as shown in Fig. 3 (a). The horizontal dimension corresponds to the depth of the NN, with each layer performing convolutional operations as follows:

$$a_{v,\ell} = F_{v,\ell}(a_{v,\ell-1}), \quad (7)$$

where $a_{v,\ell}$ represents the output feature of the $\ell$-th layer at the $v$-th scale, and $F_{v,\ell}$ denotes the convolutional operation in the $\ell$-th layer at the $v$-th scale [8]. The vertical dimension corresponds to the feature scale, which indicates the size of the feature matrix within a layer. Along this dimension, the NN employs a strided convolutional operation [28] to process features, thereby reducing the feature scale and aggregating information. It can be represented as:

$$a_{v,\ell} = \tilde{F}_{v,\ell}(a_{v-1,\ell}), \quad (8)$$

where $\tilde{F}_{v,\ell}$ denotes the strided convolution operation applied in the $\ell$-th layer at the $v$-th scale.

*3) Dense connectivity:* The dynamic TOE has dense connectivity where layers are connected to every previous layer in a feed-forward fashion. Dense connectivity can stabilize the feature propagation process as the NN deepens [28]. For scale $v \neq 1$, features from the previous scale $v-1$ are concatenated with features of the current scale, enabling multi-scale feature aggregation. It can be expressed as, for $v = 1$,

$$a_{v,\ell} = F_{v,\ell}([a_{v,\ell-1}, \ldots, a_{v,1}]). \quad (9)$$

For $v \neq 1$,

$$a_{v,\ell} = \begin{bmatrix} \tilde{F}_{v,\ell}([a_{v-1,\ell-1}, \ldots, a_{v-1,1}]) \\ F_{v,\ell}([a_{v,\ell-1}, \ldots, a_{v,1}]) \end{bmatrix}, \quad (10)$$

where $[\cdot]$ is concatenation operation.

*4) Confidence scoring:* The dynamic TOE computes a confidence score $\phi_{j,k}$ at exit $k$ for input image $s_j$ by

$$\phi_{j,k} = \frac{e^{\lambda_{j,k}^{\Theta}}}{\sum_{m=1}^{M} e^{\lambda_{j,k}^{m}}}, \quad (11)$$

JOURNAL OF LATEX CLASS FILES, VOL. 14, NO. 8, AUGUST 2021 6

where

$$\Theta = \arg\max_{m \in M} \lambda_{j,k}^m. \quad (12)$$

In the equation, $\lambda_{j,k}^m$ is the logit (raw output of NNs) calculated for sample $s_j$ at exit $k$ with predicted class label $m$ using the dynamic TOC model. $\Theta$ is the maximum logit class label of $m$ classes. Then $\sum_{m=1}^{M} e^{\lambda_{j,k}^m}$ sums up all logits over $M$ classes. A higher confidence score $\phi_{j,k}$ indicates more confidence that this image feature will be classified correctly in the later inference stage, thereby identifying it as a simple image. On the contrary, a lower confidence score suggests a more complex image. Consequently, in the validation stage, at each exit, all images are sorted in descending order based on their confidence scores $\phi_{j,k}$ to decide the exit thresholds.

*5) Computational complexity analysis:* Next, we will present the computation complexity approximation for the proposed TOE network. Given that the convolutional layers predominantly contribute to the computational cost compared to batch normalization and activation layers, we adopt the simplification approach proposed in [9], considering only the FLOPs associated with convolutional operations.

**Proposition 1.** *The computational complexity of the dynamic TOE network can be approximated by summing the total FLOPs over all layers and resolution scales. Specifically, the FLOPs for a single layer at a given scale are calculated as:*

$$\chi_{v,\ell} = 2 \cdot a_{v,\ell}^{Cin} \cdot a_{v,\ell}^{Cout} \cdot \xi^2 \cdot a_{v,\ell}^{Hout} \cdot a_{v,\ell}^{Wout}, \quad (13)$$

*where $\chi_{v,\ell}$ denotes the FLOPs at layer $\ell$ and scale $v$, $a_{v,\ell}^{Cin}$ and $a_{v,\ell}^{Cout}$ denote the numbers of input and output feature channels at layer $\ell$ and scale $v$ respectively; $\xi$ represents the convolution kernel size; and $a_{v,\ell}^{Hout}$, $a_{v,\ell}^{Wout}$ correspond to the height and width of the output feature, respectively. Consequently, the total computational complexity of the TOE network is obtained as follows:*

$$\chi = \sum_{\ell \in \mathcal{L}} \sum_{v \in \Upsilon} \chi_{v,\ell}, \quad (14)$$

Proof. The proof is provided in Appendix A.

### C. Dynamic TOC model training, validating and testing

In what follows, we will explain how dynamic TOC models are trained, validated, and tested.

*1) Dynamic TOC model training:* The loss function for training the dynamic TOC model is

$$\mathcal{L}_{dynamic} = \sum_k \omega_k \mathcal{L}_k(\mathcal{Z}, \hat{\mathcal{Z}}; \beta, \eta, \gamma, \kappa), \quad (15)$$

where $\omega_k$ is the weight of loss at the $k$-th exit (here we use $\omega_k = 1$ according to the empirical value from reference [28]), and $\mathcal{L}_k$ represents the loss function at the $k$-th exit. Similar to the static loss function in (5), $\mathcal{L}_k$ is calculated as:

$$\mathcal{L}_k(\mathcal{Z}, \hat{\mathcal{Z}}; \beta, \eta, \gamma, \kappa) = -\sum_{j=1}^{M} z_j \log(\hat{z}_{j,k}), \quad (16)$$

where $z_j$ is the true label of sample $s_j$, $\hat{z}_{j,k}$ is the predicted label of of sample $s_j$ from the $k$-th exit.

**Algorithm 1** Training process of dynamic TOC model
**Input** Training dataset $\mathcal{S}_{\text{train}}$, the number of iterations $T$, distributions of channel gain $h$ and noise $\epsilon$, batch size $B_{batch}$, number of exits $K$
**Output** The network $\varepsilon_t(\cdot)$, $\varepsilon_c(\cdot)$, $\delta_c(\cdot)$, $\delta_t(\cdot)$, FLOPs$^{\text{min}}$, FLOPs$^{\text{max}}$
**Initialization** epoch $t = 1$
1: **while** $t \leq T$ **do**
2:     Select a mini-batch of data $\mathcal{S}_{\text{train}} = \{(s_j, z_j)\}_{j=1}^{B_{batch}}$
3:     Compute task-oriented features $\varepsilon_t(\mathcal{S}; \beta) \to \mathcal{M}$
4:     Compute channel encoded features $\varepsilon_c(\mathcal{M}; \eta) \to \mathcal{X}$
5:     Transmit $\mathcal{X}$ over the channel
6:     Receive $\mathcal{Y} = h\mathcal{X} + \epsilon$
7:     Decode channel features $\delta_c(\mathcal{Y}; \kappa) \to \hat{\mathcal{M}}$
8:     Task-oriented inference $\delta_t(\hat{\mathcal{M}}; \gamma) \to \hat{\mathcal{Z}}$
9:     Compute the loss function $\mathcal{L}_{\text{dynamic}}$ by equation (15)
10:    Train $\beta, \eta, \gamma, \kappa$ using SGD optimization
11:    $t \leftarrow t + 1$
12: **end while**
13: Compute FLOPs$^{\text{min}}$ and FLOPs$^{\text{max}}$

The training process of the dynamic TOC model is shown in Algorithm 1. Before training, the number of exits $K$ for the dynamic encoder is pre-set based on device capability and processing needs. After training, the TOC model outputs the NN and related FLOPs. Due to the multiple exits in dynamic encoder NN, the expected FLOPs of the model have a range from FLOPs$^{\text{min}}$ to FLOPs$^{\text{max}}$. The FLOPs$^{\text{min}}$ is calculated when all features are output from the first exit, and the FLOPs$^{\text{max}}$ is evaluated when all features are output from the $K$-th exit. These two values indicate the lower and upper bounds of computation complexity of the dynamic TOC model.

*2) Dynamic TOC model validating:* After training the dynamic TOC model, with the output NN and related FLOPs$^{\text{min}}$ and FLOPs$^{\text{max}}$, we need to decide the exit probability $Pr_k$ at each exit $k$ to meet device budget limits. Note that for a simplifed illustration, we assume the exits are evenly distributed in the NN in terms of computation complexity, such that $\varphi$ operations (in FLOPs) are performed between an exit $k(k > 1)$ and the previous exit $k - 1$. We note that this can be implemented by e.g., choosing exits in NNs. Therefore, the FLOPs$^{\text{min}}$ is $\varphi$, and the FLOPs$^{\text{max}}$ is $K \times \varphi$, where $K$ is the total number of exits, and $K$ is a positive integer. We propose a method for distributing image samples at each exit $k$ as follows: at an exit $k(k > 0)$, the probability of an image feature to exit is given by

$$Pr_k = \frac{r^k}{\sum_{j=1}^{K} r^j}, \quad (17)$$

where the features at exit $k$ are weighted by $r^k$, normalized across all $K$ exits. $r$ is a hyperparameter that is decided based on budget $B$. For one image in a batch of $N$ images, the computation budget should be within the range $[\varphi, K\varphi]$. Under the proposed distribution method for $Pr_k$, we denote

**Algorithm 2** Validating process of dynamic TOC model
**Input:** Validation dataset $\mathcal{S}_{\text{val}}$, exit probability $Pr_k$, number of exits $K$
**Output:** Thresholds $\theta_k$
1: **for** each $s_j \in \mathcal{S}_{\text{val}}$ **do**
2:    **for** each exit $k < K$ **do**
3:       Compute confidence score $\phi_{j,k}$ for $s_j$
4:       Sort $\phi_{j,k}$ in descending order
5:    **end for**
6:    Set threshold $\theta_k$ such that the probability for $s_j$ with $\phi_{j,k}$ to exit is $Pr_k$
7: **end for**
8: Set threshold $\theta_K$ to accept all remaining samples

**Algorithm 3** Testing process of dynamic TOC model
**Input:** Test dataset $\mathcal{S}_{\text{test}}$, thresholds $\theta_k$, number of exits $K$
**Output:** Accuracy, FLOPs
1: **for** each $s_j \in \mathcal{S}_{\text{test}}$ **do**
2:    **for** each exit $k \leq K$ **do**
3:       Compute confidence score $\phi_{j,k}$ using equation (11)
4:       **if** $\phi_{j,k} \geq \theta_k$ **then**
5:          Output feature at exit $k$
6:          Transmit over the wireless channel
7:          **break**
8:       **end if**
9:    **end for**
10: **end for**
11: Compute overall accuracy and FLOPs

the expected FLOPs for one image to be $\mathcal{R}$, and it can be calculated as:

$$\mathcal{R} = \sum_{k=1}^{K} k \cdot \varphi \cdot \frac{r^k}{\sum_{j=1}^{k} r^j}. \quad (18)$$

Next, we will show that, for the proposed distribution method in (17), we can find a hyperparameter $r_{up}$ as the upper bound of $r$, such that the expected computation operations for a batch of $N$ images $N\mathcal{R}$ remain below the device computation budget $B$.

**Lemma 1**: *For any positive integer $K$ and $r \in (0, +\infty)$, the expected FLOPs $\mathcal{R}$ is monotonically increasing as a function of $r$.*

Proof. The proof is provided in Appendix B.

**Lemma 2**: *For any positive integer $K$ and $r \in (0, +\infty)$, the expected FLOPs $\mathcal{R}$ is within the range $[\varphi, K\varphi]$.*

Proof. The proof is provided in Appendix C.

**Proposition 2**: *Based on Lemma 1 and Lemma 2, there exists an $r_{up}$, such that for any $r \leq r_{up}$, the expected FLOPs for a batch of $N$ images satisfy $N\mathcal{R} \leq B$.*

Proof. The proof is given in Appendix D.

Given the exit probability $Pr_k$ of every exit, we can calculate the exit thresholds $\theta_k$ of confidence scores $\phi_{j,k}$, such that at each exit $k$, the probability for one image feature to exit is $Pr_k$. The calculation is implemented on the validation dataset $\mathcal{S}_{\text{val}}$ as shown in Algorithm 2.

*3) Dynamic TOC model testing:* With exit threshold $\theta_k$, we can apply the dynamic TOC model to the test dataset $\mathcal{S}_{\text{test}}$. The testing process of the proposed model is shown in Algorithm 3. The dynamic TOC encoder outputs features when the confidence score $\phi_{j,k}$ of the current sample $s_j$ is higher than the exit threshold $\theta_k$ at exit $k$. This feature is then passed to the wireless communication block and subsequently processed by the inference block. Finally, we calculate the overall accuracy and FLOPs.

## VI. CONVERGENCE ANALYSIS OF THE PROPOSED METHOD

In what follows, we will analyze the convergence of the proposed methods using SGD optimization under the following assumptions. Assumption 1 and Assumption 2 are regular smoothness conditions, and are approximately satisfied by most NN models, particularly when using smooth activation functions [30], [31]. Assumption 3 assumes bounded stochastic gradient noise, which typically holds under mini-batch training due to finite sample variance [31]. Note that all components in our systems have NN-based structures, including the TOE, the channel encoder, the channel decoder, and the inference block. Consequently, no convexity assumptions are made for SGD optimization problem in NNs [29].

**Assumption 1**: *The objective function of the proposed NN $f(x)$ in the training process is second differentiable*[30], [31].

**Assumption 2**: *The gradients of the objective function are Lipschitz-continuous* [31], i.e.,

$$\|\nabla^2 f(x)\| \leq L, \quad (19)$$

where $L$ is a positive constant referred to as the Lipschitz constant.

**Assumption 3**: *The noise in the gradients has bounded variance* [30], i.e.,

$$\mathbb{E}\left[\|\nabla \tilde{f}_t(x) - \nabla f(x)\|^2\right] \leq \sigma^2. \quad (20)$$

**Lemma 3**: *For the update rule of the weights in the objective function*

$$w_{t+1} = w_t - \alpha_t \nabla \tilde{f}_t(w_t), \quad (21)$$

*the objective function at the next time step satisfies:*

$$f(w_{t+1}) \leq f(w_t) - \alpha_t \nabla \tilde{f}_t(w_t)^\top \nabla f(w_t) + \frac{\alpha_t^2 L}{2} \|\nabla \tilde{f}_t(w_t)\|^2. \quad (22)$$

*Taking the expectation value, we can get*

$$\mathbb{E}[f(w_{t+1})|w_t] \leq f(w_t) - \left(\alpha_t - \frac{\alpha_t^2 L}{2}\right) \|\nabla f(w_t)\|^2 + \frac{\alpha_t^2 \sigma^2 L}{2}. \quad (23)$$

Proof. The proof is provided in Appendix E.

**Lemma 4**: *Setting step size $\alpha_t$ to be constant as $\alpha_t = \alpha$ and sum over $T$ iterations, the bound on the objective function is:*

$$\frac{\alpha}{2} \sum_{t=0}^{T-1} \mathbb{E}\left[\|\nabla f(w_t)\|^2\right] \leq f(w_0) - f^* + \frac{\alpha^2 \sigma^2 LT}{2}, \quad (24)$$

*where $f^* = \min_w f(w)$, represents the minimum value of the function $f(w)$.*

Proof. The proof is provided in Appendix F.

**Lemma 5**: Let $g_T = w_t$ with probability $\frac{1}{T}$ for all $t \in \{0, \ldots, T-1\}$. The expected value of the gradient at this point $g_T$ is

$$\mathbb{E}\left[\|\nabla f(g_T)\|^2\right] = \frac{1}{T}\sum_{t=0}^{T-1} \mathbb{E}\left[\|\nabla f(w_t)\|^2\right]. \quad (25)$$

Substituting (25) into (24), and choosing step size $\alpha = \frac{d}{\sqrt{T}}$, where $d$ is a constant value, we have

$$\mathbb{E}\left[\|\nabla f(g_T)\|^2\right] \leq \frac{1}{\sqrt{T}} \cdot \left(\frac{2(f(w_0) - f^*)}{d} + \frac{d\sigma^2 L}{2}\right). \quad (26)$$

Proof. The proof is provided in Appendix G.

**Theorem 1**: *Under Assumptions 1, 2, 3, and Lemma 5, the proposed NN-based TOC method using SGD optimization converges in expectation. Specifically, using the update rule $g_T = w_t$ with probability $\frac{1}{T}$ for all $t \in \{0, \ldots, T-1\}$, there exists a constant $D = \frac{2(f(w_0)-f^*)}{d} + \frac{d\sigma^2 L}{2} > 0$ such that*

$$\mathbb{E}[\|\nabla f(g_T)\|^2] \leq \frac{D}{\sqrt{T}}, \quad (27)$$

*and consequently, as $T \to \infty$*

$$\lim_{T \to \infty} \mathbb{E}[\|\nabla f(g_T)\|^2] = 0. \quad (28)$$

Proof. The detailed proof is provided in Appendix H.

Convergence rate measures how fast the optimization algorithm approaches a local or global minimum of the loss function during training [30]. In this context, a convergence rate of $O(1/\sqrt{T})$ means that after $T$ iterations, the expected squared gradient norm is upper bounded by a term proportional to $1/\sqrt{T}$. This convergence property implies that as training progresses, the updates become progressively smaller and more stable. Therefore, the proposed NN-based TOC method is expected to deliver stable performance after a sufficient number of training iterations.

## VII. Performance Evaluation

In what follows, we will use simulations to evaluate the performance of the proposed methods.

### A. Experimental setup

*1) Dataset:* In our simulations, we use three datasets: CIFAR-100 [32], CIFAR-10 [32], and ImageNet [7]. The CIFAR-100 dataset [32] comprises 100 distinct classes, with 50,000 training images, 10,000 testing images, and 5,000 validation images, each sized at $32 \times 32$ pixels. Similarly, the smaller-scale CIFAR-10 dataset [32] includes 10 classes and also contains 50,000 training images, 10,000 testing images, and 5,000 validation images, each image measuring $32 \times 32$ pixels. The larger-scale ImageNet dataset [7] contains approximately 1.2 million training images, 50,000 validation images, and 100,000 testing images across 1,000 classes, originally varying in dimensions but resized and cropped to $224 \times 224$ pixels for standard training protocols. For data augmentation purposes, the CIFAR-10 and CIFAR-100 datasets undergo preprocessing described in [9], including zero-padding by 4 pixels on each side, random cropping back to $32 \times 32$ pixels, horizontal flipping with a probability of 0.5, and normalization.

TABLE II
HYPERPARAMETERS IN STATIC AND DYNAMIC MODELS

| Hyperparameter | Static | Dynamic (CIFAR-10/100) | Dynamic (ImageNet) |
|---|---|---|---|
| Epochs | 164 | 300 | 90 |
| Batch size | 64 | 64 | 256 |
| Learning rate | 0.1 | 0.1 | 0.2 |
| Weight decay | 0.1 | 0.1 | 0.1 |
| Optimizer | SGD | SGD | SGD |
| Momentum | 0.9 | 0.9 | 0.9 |
| Transmission dimension | 16 | 16 | 64 |
| Number of GPUs | 1 | 1 | 8 |
| Number of Exits | - | 5 | 5 |

*2) Baselines:* We will first compare the proposed static model against three NN-based communication models: DeepJSCC [3], VFE [13] and GatedJSCC [10] models as baselines. Then we will compare the proposed dynamic model with all static models.

- DeepJSCC [3]: an NN-based communication system, where the encoder and decoder are jointly trained. It aims to reconstruct the source image after transmission. A classifier is applied to reconstructed images to evaluate model accuracy.
- VFE [13]: an NN-based TOC model. It uses VIB to select the minimum but sufficient information before and after channel transmission.
- GatedJSCC [10]: an NN-based sematic communication model. It uses coding reduction maximization to learn discriminative features. In our experiments, we only use the classification task branch.

*3) Performance metrics:* Four performance metrics are used in our experiments.

- Transmitted feature dimension: This metric quantifies the feature dimension transmitted through wireless transmission channels. Fewer transmitted dimensions indicate a smaller transmission load.
- FLOPs: FLOPs measure the network computation complexity. It is calculated based on the input feature size and the structure of NN (e.g., number of layers, layer types).
- Accuracy at different PSNR levels: Accuracy in the classification problem is defined as the number of correct inference results divided by the total number of inference results. PSNR used in the experiments is defined as

$$\text{PSNR} = 10 \log_{10}\left(\frac{P}{\sigma^2}\right) \text{ (dB)}, \quad (29)$$

where $P$ is the signal power, and $\sigma^2$ is the power of channel noise. Accuracy at different PSNR levels indicates the robustness of the TOC systems.
- Accuracy-FLOPs curve: The metric calculates model accuracy at different FLOPs levels.

*4) Hyperparameters:* The hyperparameters employed in the experiments are summarized in Table II, and the selection process is detailed in Appendix I.

### B. Static model results

In what follows, we will compare the performance of the proposed static model with DeepJSCC, VFE, and GatedJSCC

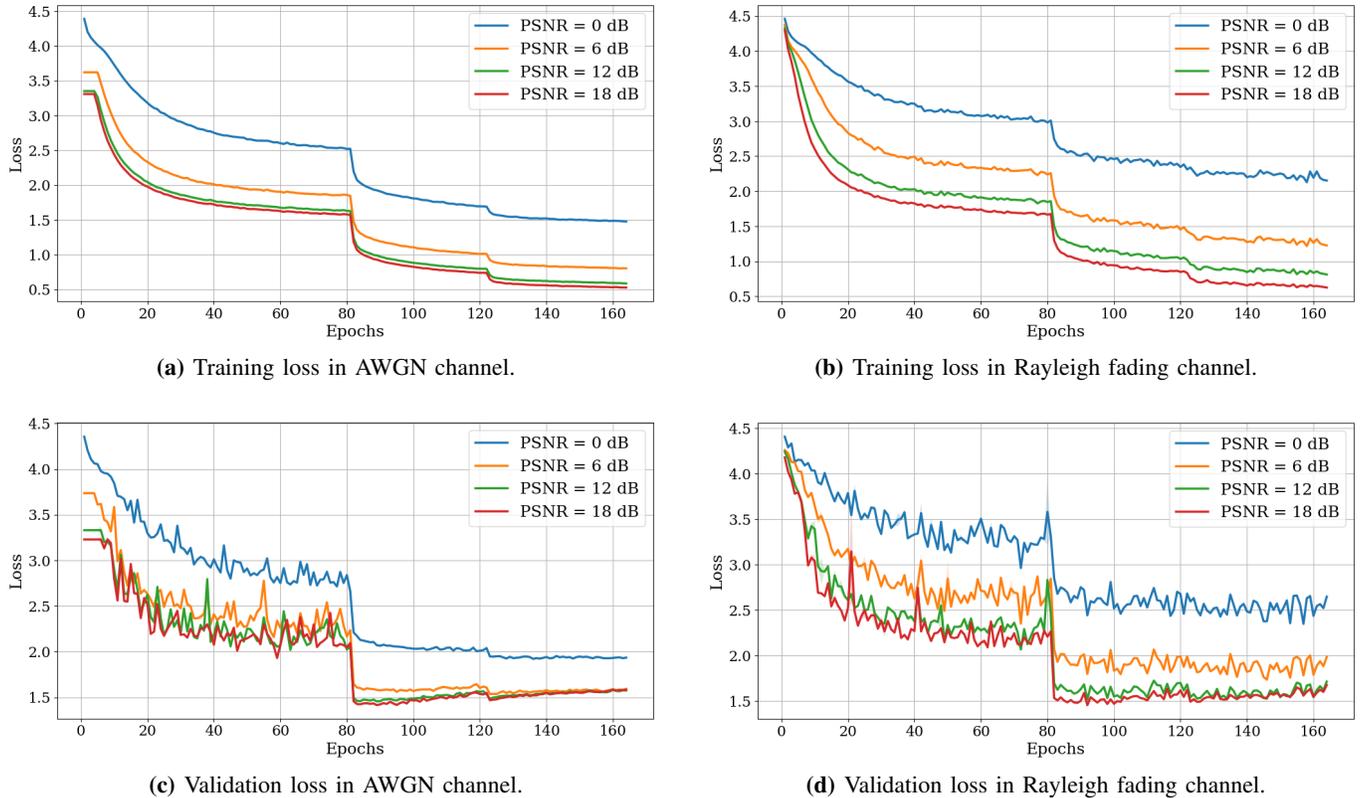

**(a)** Training loss in AWGN channel.

**(b)** Training loss in Rayleigh fading channel.

**(c)** Validation loss in AWGN channel.

**(d)** Validation loss in Rayleigh fading channel.

Fig. 4: The training and validation loss of the proposed static model in AWGN and Rayleigh channels in CIFAR-100 dataset.

models. The results are presented in Table III. We utilize different sizes of ResNet [9] as the static TOE and evaluate their performance in different PSNR values in both AWGN and Rayleigh fading channels. We also add one attention-based model ViT-Tiny [33] as the static semantic encoder for comparison. All results are averaged over 5 experiment trials.

*1) Feature dimensions, FLOPs, and accuracy:* As shown in Table III, the proposed static TOC model transmits fewer feature dimensions compared to the three baseline models. At the proposed channel encoder, the task features are compressed to 16 dimensions. In contrast, DeepJSCC model requires a full reconstruction of the original input image after channel transmission, therefore using 2048 dimensions. VFE model adapts its transmitted dimension depending on channel conditions. It sets the maximum transmission limit of 64 dimensions and uses $21 - 26$ dimensions in the experiments. GatedJSCC model requires 256 dimensions for channel transmission. The proposed static TOC model achieves a $99.22\%$ reduction in transmitted dimensions compared to DeepJSCC model, $23.81\%$ reduction compared to VFE model and $93.75\%$ reduction compared to GatedJSCC model.

Furthermore, the proposed static TOC model exhibits reduced computational complexity in terms of FLOPs. With ResNet-20 as the TOE, the static model achieves approximately $40.81$ million FLOPs, resulting in reductions of $18.79\%$ compared to DeepJSCC ($50.25$ million FLOPs) and $89.13\%$ compared to VFE ($380.55$ million FLOPs). Compared to GatedJSCC, the FLOPs are comparable due to the simple network architecture employed in GatedJSCC model. These results highlight the benefits of the proposed approach in reducing computational complexity, leading to faster training and inference. Additionally, as deeper networks are adopted for the static TOE (e.g., transitioning from ResNet-20 to ResNet-110), the FLOPs proportionally increase, indicating a higher demand for computational resources. Meanwhile, the ViT-Tiny based model incurs significantly higher complexity ($366.68$M FLOPs) than ResNet-based models, due to the overhead introduced by its self-attention mechanisms [33].

The accuracy of the proposed static TOC model at different PSNR levels in both AWGN and Rayleigh fading channels is shown in Table III. Note that VFE model does not provide a way to learn features in Rayleigh fading channels [13]. With ResNet-20 as TOE, the proposed static model outperforms both DeepJSCC, VFE and GatedJSCC models in various PSNR levels in terms of accuracy, especially at high PSNR values. For example, in AWGN channel at PSNR = 0 dB, the static model with ResNet-20 achieves an accuracy of $55.58\%$, outperforming the JSCC model by $41.75\%$, VFE model by $35.57\%$, and GatedJSCC by $6.70\%$. The accuracy fluctuates as the TOE network grows deeper from ResNet-20 to ResNet-110. This can be explained by the fact that a deeper network has a higher chance of overfitting the training dataset and performing worse on the testing dataset [34]. In contrast, the ViT-Tiny model used as a semantic encoder performs worse than ResNet-based models. This is due to attention-based architectures like ViT requiring extensive pretraining

TABLE III
COMPARISON OF STATIC MODELS IN AWGN AND RAYLEIGH FADING CHANNELS.

| Category | Model | Feature Dimension | FLOPs | Accuracy (%) (AWGN) | | | | Accuracy (%) (Rayleigh fading) | | | |
|---|---|---|---|---|---|---|---|---|---|---|---|
| | | | | 0dB | 6dB | 12dB | 18dB | 0dB | 6dB | 12dB | 18dB |
| **Baseline** | JSCC | 2048 | 50.25M | 39.22 | 48.44 | 51.14 | 52.33 | 36.20 | 46.02 | 48.86 | 53.90 |
| | VFE | 21–26, up to 64 | 380.55M | 41.00 | 56.62 | 59.41 | 59.46 | - | - | - | - |
| | GatedJSCC | 256 | blue40.52M | 52.14 | 52.91 | 54.37 | 54.73 | 53.28 | 53.30 | 53.53 | 54.39 |
| **Proposed** | Static_ResNet20 | 16 | **40.81M** | **55.58** | **62.44** | **62.03** | **62.32** | 34.98 | 54.16 | 61.07 | 63.78 |
| | Static_ResNet32 | 16 | 69.24M | 64.94 | 63.68 | 65.16 | 66.32 | 34.57 | 48.92 | 58.81 | 61.34 |
| | Static_ResNet44 | 16 | 97.66M | 54.32 | 56.85 | 61.87 | 65.08 | 39.19 | 52.97 | 56.75 | 63.53 |
| | Static_ResNet56 | 16 | 126.09M | 59.85 | 57.95 | 65.62 | 64.94 | 39.87 | 47.20 | 58.85 | 63.18 |
| | Static_ResNet110 | 16 | 254.01M | 45.54 | 62.57 | 65.43 | 65.55 | 37.41 | 53.21 | 59.63 | 65.97 |
| | Static_ViT_Tiny | 16 | 366.68M | 48.25 | 48.85 | 49.01 | 50.09 | 45.99 | 49.03 | 49.53 | 51.00 |

TABLE IV
STATIC AND DYNAMIC MODEL ACCCURACY WITH DIFFERENT BUDGET $B$ AT PSNR = 0 DB IN AWGN CHANNELS.

| | Model | budget $B$ (FLOPs) | | | | | | | | | |
|---|---|---|---|---|---|---|---|---|---|---|---|
| | | 10M | 20M | 30M | 40M | 50M | 60M | 70M | 100M | 130M | 260M | 380M |
| Baseline Accuracy (%) | JSCC | - | - | - | - | - | 39.22 | 39.22 | 39.22 | 39.22 | 39.22 | 39.22 |
| | VFE | - | - | - | - | - | - | - | - | - | - | 41.00 |
| Static Model Accuracy (%) | Static + ResNet20 | - | - | - | - | 55.58 | 55.58 | 55.58 | 55.58 | 55.58 | 55.58 | 55.58 |
| | Static + ResNet32 | - | - | - | - | - | - | 64.94 | 64.94 | 64.94 | 64.94 | 64.94 |
| | Static + ResNet44 | - | - | - | - | - | - | - | 54.32 | 54.32 | 54.32 | 54.32 |
| | Static + ResNet56 | - | - | - | - | - | - | - | - | 59.85 | 59.85 | 59.85 |
| | Static + ResNet110 | - | - | - | - | - | - | - | - | - | 45.54 | 45.54 |
| Dynamic Model Accuracy (%) | Dynamic | **51.41** | **57.86** | **61.26** | **62.03** | **62.26** | **62.26** | **62.26** | **62.26** | **62.26** | **62.26** | **62.26** |

on large-scale datasets [35], making them less effective for smaller datasets such as CIFAR-100.

*2) Static model convergence:* We use the static model with ResNet-20 as an example to show the model training and validation convergence. We show the training loss of the static model for AWGN and Rayleigh fading channels in Fig. 4a and Fig. 4b, and the validation loss in Fig. 4c and Fig. 4d. The figures demonstrate that training and validation loss decreases over epochs, eventually converging to a stable value. The results are consistent with our convergence analysis in Section VI. Additionally, the loss function achieves lower values at high PSNR levels and higher values at low PSNR levels. This suggests that the system is less stable and harder to converge in more noisy channels.

### C. Dynamic model results

In dynamic TOC models, we will first give the input image sorting result. Then, we will give the model accuracy based on different device budgets and the accuracy-FLOPs curves. We will also present the training and validation loss to demonstrate the convergence of the proposed dynamic TOC model.

*1) Dynamic TOE sorting results:* Fig. 5 gives an example of how the dynamic TOE outputs simple image features from early exits, and outputs complex image features from late exits. This is measured at PSNR = 6 dB in Rayleigh fading channel, with exits number $K = 5$. We set the exit probability $Pr_k$ at every exit to be equal, where $Pr_k = 1/K$. As shown in the figure, the images from the early exits are characterized by clear and distinctive features, which are considered simple to classify. In contrast, images from the late exits typically

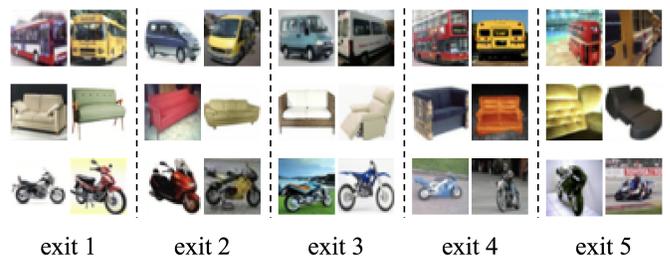

exit 1    exit 2    exit 3    exit 4    exit 5

Fig. 5: Dynamic model outputs at PSNR = 6 dB in Rayleigh fading channels, displaying example images leaving from exit 1 to 5 in CIFAR-100 dataset.

display sideways angles, or atypical colors and shapes, which are considered complex to classify.

*2) Model accuracy for different computation budget:* We will present the model accuracy at different computation budgets, using an example result in AWGN channel at PSNR = 0 dB. The accuracy of the two baseline models, the proposed static model and the dynamic model are shown in Table IV. In the table, a "−" sign indicates that the model cannot be used under the current device budget. The table shows that our proposed static model is more flexible, supporting different device budgets when using different static TOE networks. However, as the budget continually increases, neither the baseline models nor the proposed static TOC model show an increase in accuracy. Conversely, the proposed dynamic model not only supports different budgets but also demonstrates increasing accuracy with higher budgets. This is particularly

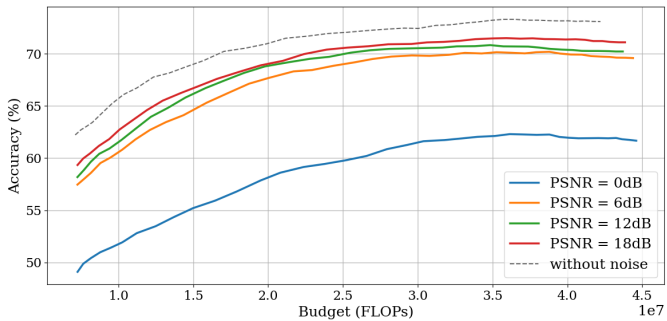

**(a)** AWGN channel.

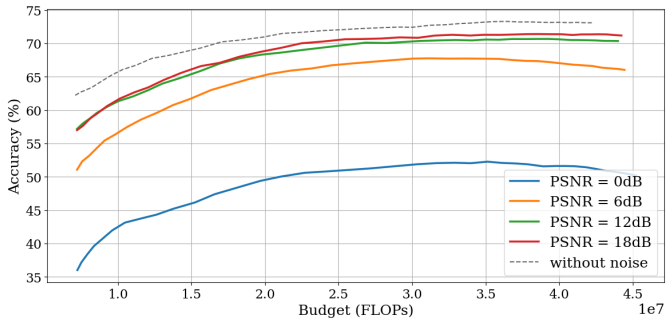

**(b)** Rayleigh fading channel.

Fig. 6: Accuracy as a function of budget $B$ in AWGN and Rayleigh fading channels in CIFAR-100 dataset.

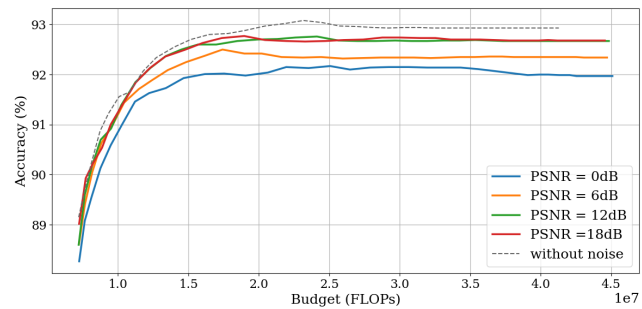

Fig. 7: Accuracy as a function of budget $B$ in AWGN channel on CIFAR-10 dataset.

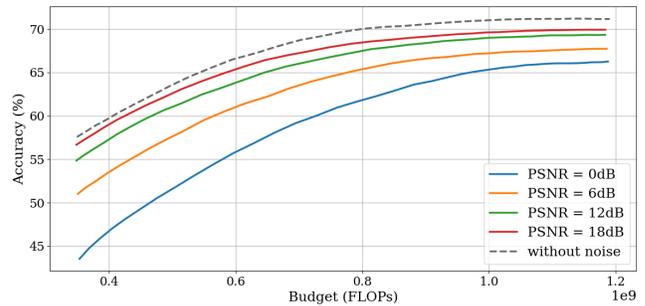

Fig. 8: Accuracy as a function of budget $B$ in Rayleigh fading channel on ImageNet dataset.

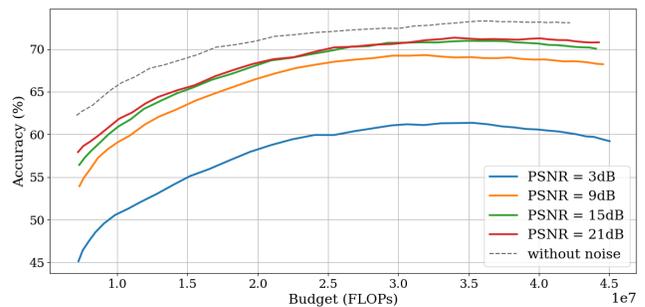

Fig. 9: Accuracy as a function of budget $B$ in Rayleigh fading channel on CIFAR-100 dataset (more PSNR values).

useful when devices with different computation budgets are used in TOC. Additionally, it is observed that the highest accuracy of the proposed dynamic model also outperforms the two baseline models in terms of accuracy, achieving a 58.75% increase compared to DeepJSCC model and a 51.83% increase compared to VFE model.

*3) Accuracy-FLOP:* In what follows, we will present the curve of accuracy as a function of budget $B$ across different PSNR values. The results for AWGN channels are shown in Fig. 6a and for Rayleigh fading channels in Fig. 6b on the CIFAR-100 dataset. We also provide dynamic model results on CIFAR-10 dataset in Fig. 7, and dynamic model results on ImageNet dataset in Fig. 8. The dashed line in the figure indicates model performance without channel noise, representing the ideal case and serving as an upper bound for comparison. These results collectively demonstrate the effectiveness of our method across both small-scale and large-scale datasets. It is observed that at higher PSNR values, the proposed dynamic model achieves higher accuracy. The curve of PSNR = 0 dB shows significant fluctuations, indicating that the model is substantially influenced by channel noise. Generally, accuracy increases with a higher device budget $B$. However, the highest accuracy point is not necessarily achieved at the highest budget $B$, where the model attains its best performance with lower FLOPs, can be identified. This may be explained that simpler images are more effectively classified at earlier exit points, while deeper layers may introduce overfitting and lead to misclassification [34]. Therefore, the optimal configuration depends on both the input data complexity and model behavior, suggesting that practical deployments can benefit from dynamic strategies that adaptively allocate computational resources. To further validate the robustness of the proposed dynamic model, we include more PSNR values results in Fig. 9.

*4) Dynamic model convergence:* Then we will present the training and validating loss of the proposed dynamic TOC model. The training loss curve in AWGN channels and Rayleigh fading channels is shown in Fig. 10a and Fig. 10b, respectively, and the validation loss in AWGN channels and Rayleigh fading channels in Fig. 10c and Fig. 10d, respectively. Consistent with the static models, the loss in dynamic models stabilizes over iterations, indicating model convergence. Moreover, the results show a lower loss for higher PSNR levels.

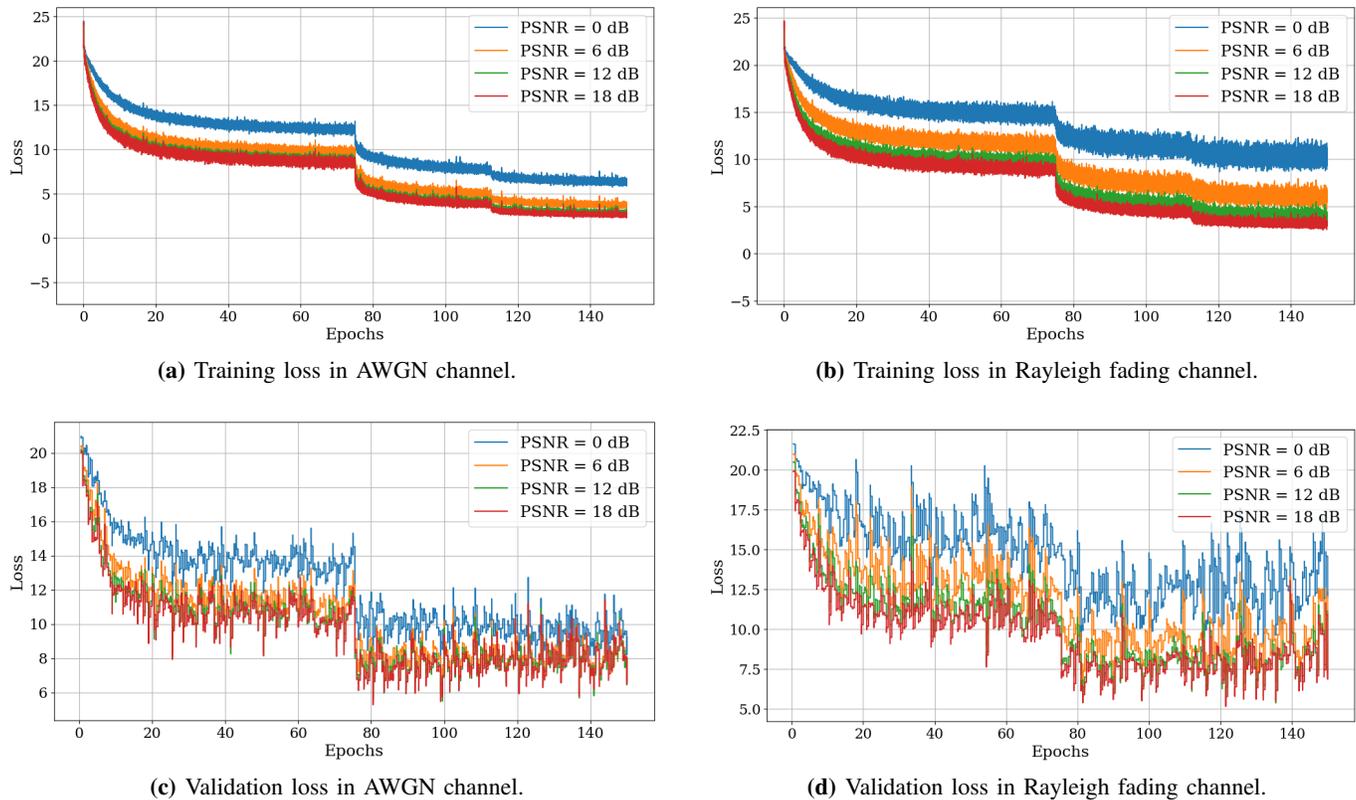

Fig. 10: The training and validation loss for AWGN and Rayleigh fading channels in the dynamic TOC model on the CIFAR-100 dataset. All models converge to a stable point.

## VIII. CONCLUSIONS

In this paper, we have proposed computation-efficient task-oriented communication (TOC) methods with static and dynamic models. These models consist of task configuration, task-oriented encoding, wireless communication, and task execution blocks. Our wireless transmission block incorporates a simplified NN-based channel encoder and decoder, which reduces the transmitted feature dimensions and computation complexity. The static TOE uses ResNet to extract task-related features, enhancing feature extraction ability and flexibility. The proposed static TOC model achieves higher accuracy with lower transmitted dimensions and computation resources than baseline models (DeepJSCC, VFE, and GatedJSCC models). On top of the benefits of the static TOC model, the proposed dynamic model further improves performance when devices have limited computational budgets and tasks have varying complexity. For this purpose, the dynamic model uses a horizontal and vertical dimension multi-exit NN structure that calculates the confidence score of every input image and allocates computational resources based on input image complexity. The experiment results show that the dynamic model can accommodate various computation budgets, and the accuracy grows as the budget increases. We also provide theoretical proof of the convergence for NN-based TOC models using SGD optimization. Experimental results further validate the model convergence.

Potential applications include deploying the proposed static and dynamic models in smart industrial settings, where high-performance GPUs can exploit the capabilities of the static model, while lightweight devices like UAVs can benefit from the efficient resource allocation of the dynamic models. Nonetheless, practical implementation faces challenges such as real-time adaptation to fluctuating channel conditions and hardware constraints. Future work may focus on improving robustness under dynamic wireless environments, leveraging federated learning for distributed training, and extending the models to handle multi-modal communication tasks. Incorporating multi-modal learning and information fusion techniques could further enhance performance by leveraging diverse data types, thereby making TOC more adaptable to complex real-world scenarios. Addressing these challenges will be key to advancing the practical utility of TOC in modern wireless communication systems.

## APPENDIX A
## PROOF OF PROPOSITION 1

We provide here a method for numerically estimating the computational complexity of the dynamic TOE network. As illustrated in Fig.3 (a), the network is composed of an initial layer followed by multiple subsequent layers. The initial layer of the multi-exit dynamic TOE consists of multiple resolution scales, denoted by $v$. Each scale $v \in \Upsilon$ includes a convolutional layer, a batch normalization layer, a ReLU activation, and a pooling layer. Since the computational cost of convolutional and linear layers significantly outweighs that of batch normalization and activation layers, we follow the simplification approach in [9] and only account for FLOPs associated with the convolutional operations in our estimation.

Let $\chi_{v,\ell}$ denote the FLOPs required for layer $\ell$ at scale $v$. It is computed as:

$$\chi_{v,\ell} = 2 \cdot a_{v,\ell}^{\text{Cin}} \cdot a_{v,\ell}^{\text{Cout}} \cdot \xi^2 \cdot a_{v,\ell}^{\text{Hout}} \cdot a_{v,\ell}^{\text{Wout}}, \tag{30}$$

where $a_{v,\ell}^{\text{Cin}}$ and $a_{v,\ell}^{\text{Cout}}$ denote the number of input and output feature channels, respectively, in layer $\ell$ at scale $v$. $\xi$ represents the kernel size, and $a_{v,\ell}^{\text{Hout}}$, $a_{v,\ell}^{\text{Wout}}$ correspond to the height and width of the output feature. The channel dimensions and spatial resolutions at each scale vary according to scale-dependent hyperparameters $\tau_v$ and $\iota_v$, with the general trend that channel dimensions increase and spatial dimensions decrease as $v$ increases. This relationship can be modeled as:

$$a_{v,\ell}^{\text{Cin}} = \tau_v \cdot a_{v-1,\ell-1}^{\text{Cin}}, \tag{31}$$
$$a_{v,\ell}^{\text{Cout}} = \tau_v \cdot a_{v-1,\ell-1}^{\text{Cout}}, \tag{32}$$
$$a_{v,\ell}^{\text{Hout}} = \iota_v \cdot a_{v-1,\ell-1}^{\text{Hout}}, \tag{33}$$
$$a_{v,\ell}^{\text{Wout}} = \iota_v \cdot a_{v-1,\ell-1}^{\text{Wout}}. \tag{34}$$

For the first layer, the total FLOPs is obtained by summing over all scales:

$$\chi_1 = \sum_{v \in \Upsilon} \chi_{v,1}. \tag{35}$$

For subsequent layers ($\ell > 1$), a bottleneck [9] structure is adopted. Let $\varsigma$ denote the bottleneck width factor. The effective number of input channels is adjusted as:

$$a_{v,\ell}^{'\text{Cin}} = \min\left\{a_{v,\ell}^{\text{Cin}}, \varsigma_v \cdot a_{v,\ell}^{\text{Cout}}\right\}. \tag{36}$$

The FLOPs at scale $v$ and layer $\ell > 1$ is then given by:

$$\chi_{v,\ell} = 2 \cdot a_{v,\ell}^{'\text{Cin}} \cdot a_{v,\ell}^{\text{Cout}} \cdot \xi^2 \cdot a_{v,\ell}^{\text{Hout}} \cdot a_{v,\ell}^{\text{Wout}}. \tag{37}$$

The scaling behavior of the spatial dimensions and channels in these layers follows the same formulation as in the first layer. Finally, the total FLOPs of the dynamic TOE network is computed by summing over all layers $\ell \in \mathcal{L}$ and scales $v \in \Upsilon$:

$$\chi = \sum_{\ell \in \mathcal{L}} \sum_{v \in \Upsilon} \chi_{v,\ell}. \tag{38}$$

Note that this estimation serves as a simplified analytical approximation of the network complexity. In practice, during our experiments, we compute the FLOPs in the network by accounting for all layer types, including batch normalization, activation, and pooling layers.

## APPENDIX B
## PROOF OF LEMMA 1

The expected FLOPs of the NN are:

$$\mathcal{R} = \sum_{k=1}^{K} k \cdot \varphi \cdot \frac{r^k}{\sum_{j=1}^{k} r^j} \tag{39}$$

$$= \sum_{k=1}^{K} k \cdot \varphi \cdot \frac{r^k}{\frac{r(1-r^K)}{1-r}}$$

$$= \sum_{k=1}^{K} k \cdot \varphi \cdot \frac{r^{k-1} \cdot (1-r)}{1 - r^K}$$

$$= \varphi \cdot \frac{1-r}{1-r^K} \sum_{k=1}^{K} k \cdot r^{k-1}$$

$$= \varphi \cdot \frac{1-r}{1-r^K} \cdot \frac{1 - (K+1) \cdot r^K + K \cdot r^{K+1}}{(1-r)^2}$$

$$= \varphi \cdot \frac{1 - (K+1) \cdot r^K + K \cdot r^{K+1}}{(1-r) \cdot (1-r^K)}.$$

$$\frac{\partial \mathcal{E}}{\partial r} = \frac{2(K^2-1) \cdot r^{K+1} - K^2 r^{K+2} - K^2 r^K + r^{2K+1} + r}{(1-r)^2 \, r \, (r^K - 1)^2}. \tag{40}$$

For $r \neq 1$, the denominator $(1-r)^2 \, r \, (r^K - 1)^2 > 0$. For the nominator,

$$2(K^2-1)r^{K+1} - K^2 r^{K+2} - K^2 r^K + r^{2K+1} + r \tag{41}$$
$$= 2K^2 r^{K+1} - K^2 r^{K+2} - K^2 r^K - 2r^{K+1} + r^{2K+1} + r$$
$$= K^2 r^K (2r - r^2 - 1) + r(r^{2K} - 2r^K + 1)$$
$$= -K^2 r^K (r-1)^2 + r(r^K - 1)^2.$$

Using the arithmetic mean-geometric mean inequality (AM-GM Inequality),

$$\frac{\sum_{i=0}^{K-1} r^{-i}}{K} \geq \sqrt[K]{\prod_{i=0}^{K-1} r^{-i}}, \tag{42}$$

$$\frac{\sum_{i=0}^{K-1} r^i}{K} \geq \sqrt[K]{\prod_{i=0}^{K-1} r^i}. \tag{43}$$

$$\left(\sum_{i=0}^{K-1} r^i\right)\left(\sum_{i=0}^{K-1} r^{-i}\right) \geq K^2 \sqrt[K]{\prod_{i=0}^{K-1} r^i} \cdot \sqrt[K]{\prod_{i=0}^{K-1} r^{-i}} \tag{44}$$
$$= K^2.$$

Since

$$\frac{r^K - 1}{r - 1} = \sum_{i=0}^{K-1} r^i, \tag{45}$$

$$\frac{r^{1-K} \left(r^K - 1\right)^2}{(r-1)^2} = r^{1-K} \left(\sum_{i=0}^{K-1} r^i\right)^2 \tag{46}$$
$$= \left(\sum_{i=0}^{K-1} r^{-i}\right)\left(\sum_{i=0}^{K-1} r^i\right) \geq K^2.$$

We can prove that

$$-K^2 r^K (r-1)^2 + r(r^K - 1)^2 > 0. \tag{47}$$

Therefore, the derivative of expected FLOPs $\mathcal{E}$ w.r.t. $r$ is non-negative. The expected FLOPs as a function of $r$ are monotonically increasing.

## APPENDIX C
## PROOF OF LEMMA 2

The expected FLOPs of the NN are:

$$\mathcal{R} = \varphi \cdot \frac{1 - (K+1) \cdot r^K + K \cdot r^{K+1}}{(1-r) \cdot (1-r^K)}. \tag{48}$$

As $r \to 0$, $r^K$ and $r^{K+1}$ are very small (approaching zero). In this case, the function $f(r)$ simplifies to:

$$\mathcal{R} \approx \varphi \cdot \frac{1}{1-r} = \varphi. \tag{49}$$

As $r \to \infty$,

$$\mathcal{R} = \varphi \cdot \frac{1 - (K+1)r^k + Kr \cdot r^k}{(r^k - 1)(r - 1)} \tag{50}$$

$$= \varphi \cdot \frac{1 - r^k - K(1-r)r^k}{(r^k - 1)(r - 1)}$$

$$= \varphi \cdot \frac{1 - r^k}{1 - r^k} \cdot \frac{1}{r - 1} + \varphi \cdot \frac{K(r-1)r^k}{(r^k - 1)(r - 1)}$$

$$= \varphi \cdot \frac{1}{1 - r^k} + \varphi \cdot \frac{Kr^k}{r^k - 1}$$

$$\approx 0 + \varphi \cdot K$$

$$= K\varphi.$$

Therefore we can prove that the expected FLOPs of the dynamic NN lie within the range $[\varphi, K\varphi]$.

## APPENDIX D
## PROOF OF PROPOSITION 2

Function $\mathcal{R}(r) : r \to \mathcal{R}$ is bijective. Hence, $\mathcal{R}(r)^{-1} : \mathcal{R} \to r$ exists. Based on Lemma 1, $\mathcal{R}(r)$ is monotonically increasing. Based on Lemma 2, the expected FLOPs $\mathcal{R}$ is within the range $[\varphi, K\varphi]$, where the range for computation budget $\frac{B}{N}$ is also $[\varphi, K\varphi]$. By letting $r \leq r_{up} = \mathcal{R}(r)^{-1}(\frac{B}{N})$, hyper-parameter $r_{up}$ is found such that the expected FLOPs $\mathcal{R} \leq \frac{B}{N}$. Therefore, for a batch of $N$ images, the expected FLOPs $N \times \mathcal{R} \leq B$.

## APPENDIX E
## PROOF OF LEMMA 3

Starting with the update rule in (21), at the next time step, by Taylor's theorem, the objective will be:

$$f(w_{t+1}) = f(w_t - \alpha_t \nabla \tilde{f}_t(w_t)) \tag{51}$$

$$= f(w_t) - \alpha_t \nabla \tilde{f}_t(w_t)^\top \nabla f(w_t)$$

$$+ \frac{\alpha_t^2}{2} \nabla \tilde{f}_t(w_t)^\top \nabla^2 f(y_t) \nabla \tilde{f}_t(w_t)$$

$$\leq f(w_t) - \alpha_t \nabla \tilde{f}_t(w_t)^\top \nabla f(w_t) + \frac{\alpha_t^2 L}{2} \|\nabla \tilde{f}_t(w_t)\|^2.$$

Taking the expected value:

$$\mathbb{E}[f(w_{t+1})|w_t] \leq f(w_t) - \alpha_t \mathbb{E}\left[\nabla \tilde{f}_t(w_t)^\top \nabla f(w_t)|w_t\right] \tag{52}$$

$$+ \frac{\alpha_t^2 L}{2} \mathbb{E}\left[\|\nabla \tilde{f}_t(w_t)\|^2 | w_t\right]$$

$$= f(w_t) - \alpha_t \|\nabla f(w_t)\|^2 + \frac{\alpha_t^2 L}{2} \mathbb{E}\left[\|\nabla \tilde{f}_t(w_t)\|^2 | w_t\right]$$

$$= f(w_t) - \alpha_t \|\nabla f(w_t)\|^2 + \frac{\alpha_t^2 L}{2} \|\nabla f(w_t)\|^2$$

$$+ \frac{\alpha_t^2 L}{2} \mathbb{E}\left[\|\nabla \tilde{f}_t(w_t) - \nabla f(w_t)\|^2 | w_t\right]$$

$$\leq f(w_t) - \left(\alpha_t - \frac{\alpha_t^2 L}{2}\right) \|\nabla f(w_t)\|^2 + \frac{\alpha_t^2 \sigma^2 L}{2}.$$

## APPENDIX F
## PROOF OF LEMMA 4

If we set $\alpha_t$ small enough that $1 - \alpha_t L/2 > 1/2$, then:

$$\mathbb{E}[f(w_{t+1})|w_t] \leq f(w_t) - \frac{\alpha_t}{2} \|\nabla f(w_t)\|^2 + \frac{\alpha_t^2 \sigma^2 L}{2}. \tag{53}$$

Now taking the full expectation, summing up over an epoch of length $T$ and setting the step size to be constant $\alpha_t = \alpha$:

$$\mathbb{E}[f(w_T)] \leq f(w_0) - \sum_{t=0}^{T-1} \frac{\alpha}{2} \mathbb{E}[\|\nabla f(w_t)\|^2] + \frac{\alpha^2 \sigma^2 LT}{2}. \tag{54}$$

Rearranging the terms:

$$\frac{\alpha}{2} \sum_{t=0}^{T-1} \mathbb{E}\left[\|\nabla f(w_t)\|^2\right] \leq f(w_0) - \mathbb{E}[f(w_T)] + \frac{\alpha^2 \sigma^2 LT}{2} \tag{55}$$

$$\leq f(w_0) - \left(\min_w f(w)\right) + \frac{\alpha^2 \sigma^2 LT}{2}$$

$$\leq f(w_0) - f^* + \frac{\alpha^2 \sigma^2 LT}{2}.$$

## APPENDIX G
## PROOF OF LEMMA 5

Let $g_T = w_t$ with probability $\frac{1}{T}$ for all $t \in \{0, \dots, T-1\}$:

$$\mathbb{E}\left[\|\nabla f(g_T)\|^2\right] = \frac{1}{T} \sum_{t=0}^{T-1} \mathbb{E}\left[\|\nabla f(w_t)\|^2\right]. \tag{56}$$

Substituting back into the previous bound,

$$\frac{\alpha T}{2} \mathbb{E}\left[\|\nabla f(g_T)\|^2\right] \leq f(w_0) - f^* + \frac{\alpha^2 \sigma^2 LT}{2}. \tag{57}$$

$$\mathbb{E}\left[\|\nabla f(g_T)\|^2\right] \leq \frac{2(f(w_0) - f^*)}{\alpha T} + \frac{\alpha \sigma^2 L}{2}.$$

Now, we know that we are going to run $T$ iterations, and we can set the step size to minimize this expression. We can choose step size $\alpha = \frac{d}{\sqrt{T}}$, $d$ is a constant value, we have

$$\mathbb{E}\left[\|\nabla f(g_T)\|^2\right] \leq \frac{1}{\sqrt{T}} \cdot \left(\frac{2(f(w_0) - f^*)}{d} + \frac{d\sigma^2 L}{2}\right). \tag{58}$$

## APPENDIX H
## PROOF OF THEOREM 1

The term $\frac{1}{\sqrt{T}} \cdot \left(\frac{2(f(w_0) - f^*)}{d} + \frac{d\sigma^2 L}{2}\right)$ is a constant value $D$. Therefore, we have

$$\mathbb{E}[\|\nabla f(g_T)\|^2] \leq \frac{D}{\sqrt{T}}, \tag{59}$$

and as $T \to \infty$,

$$\lim_{T \to \infty} \mathbb{E}[\|\nabla f(g_T)\|^2] = 0. \tag{60}$$

## APPENDIX I
## HYPERPARAMETER SELECTION

### A. Ablation study on transmitted dimension

In our proposed model, for the CIFAR-10 and CIFAR-100 datasets, the channel dimension is 16. To explore its impact, we use a PSNR of 12 dB in a Rayleigh fading channel as an example and evaluate the performance at varying channel dimensions (8, 16, 32, and 64), as illustrated in Figure 11. From the results, we observe that increasing the channel dimension generally enhances the task accuracy. Specifically, increasing from 8 to 16 dimensions significantly improves accuracy. However, further increasing the dimension beyond 16 to 32 or 64 results in marginal improvements, which do not justify the additional channel transmission resource consumption.

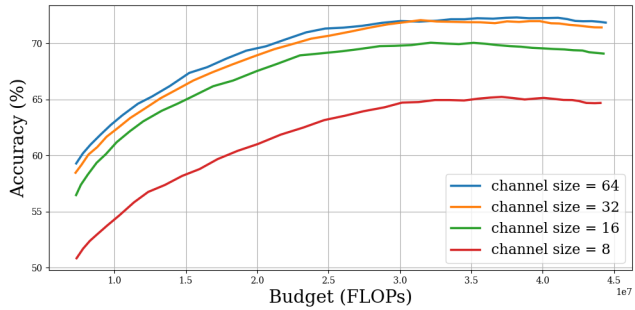

Fig. 11: Model accuracy under different channel dimensions in Rayleigh channel at PSNR = 12dB on CIFAR-100 dataset.

*B. Ablation study on learning rate*

We illustrate our learning-rate selection using a PSNR of 12 dB over a Rayleigh fading channel on the CIFAR-100 dataset as an example. The experiment employs a step-decay schedule over 300 epochs, reducing the learning rate by a factor of 0.1 at epochs 100, 200, and 300. To assess sensitivity to the initial rate, we compare five values: 0.01, 0.05, 0.1, 0.2, and 0.3, under identical training settings. As shown in Fig. 12a, higher initial rates (0.2 and 0.3) induce oscillations and plateaus in the training loss, with 0.3 failing to drop below a loss of 15 even after all decay steps. Learning rates of 0.05 and 0.1 yield smooth, monotonic loss reduction, and 0.01 converges more slowly. Correspondingly, Fig. 12b demonstrates that initial rates of 0.2 and 0.3 lead to test accuracy below 65% and 37%, respectively, while both 0.05 and 0.1 attain approximately 70% accuracy by training end. Although the learning rate of 0.05 exhibits marginally gentler convergence, a learning rate of 0.1 consistently achieves the highest final accuracy. Therefore, we recommend an initial learning rate of 0.1 with a step-decay factor of 0.1 as the balanced choice for stable convergence and optimal model performance.

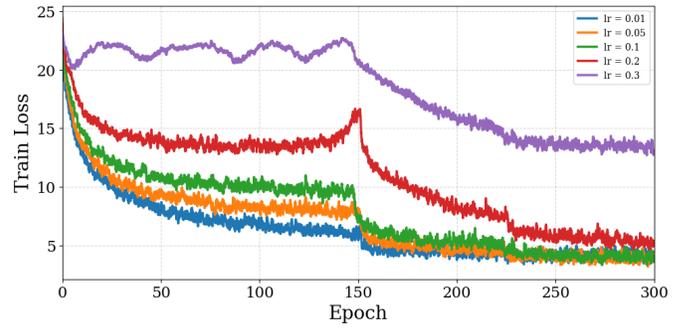

(a) Training loss at different learning rates.

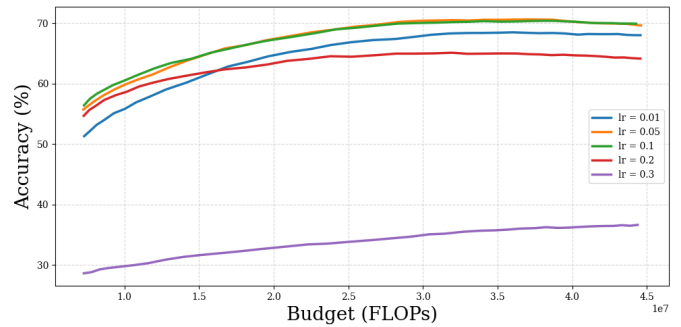

(b) Model performance at different learning rates.

Fig. 12: Ablation study on learning rate in Rayleigh channel at PSNR = 12dB on CIFAR-100 dataset.

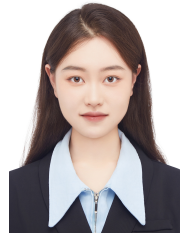

**Jingwen Fu** received the B.S. degree from Beihang University, Beijing, China, and the M.S. degree from KTH Royal Institute of Technology, Stockholm, Sweden. She is currently pursuing the Ph.D. degree with the School of Electrical Engineering and Computer Science, KTH Royal Institute of Technology, under the supervision of Prof. Ming Xiao. Her research interests include semantic communications, wireless communications, and machine learning for next-generation wireless networks.

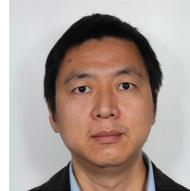

**Ming Xiao** (Senior Member, IEEE) received the bachelor's and master's degrees in engineering from the University of Electronic Science and Technology of China, Chengdu, in 1997 and 2002, respectively, and the Ph.D. degree from the Chalmers University of Technology, Sweden, in November 2007. Since November 2007, he has been with the Department of Information Science and Engineering, School of Electrical Engineering and Computer Science, KTH Royal Institute of Technology, Sweden, where he is currently a Professor. He received the IEEE Vehicular Technology Society Best Magazine Paper Award 2023. He was an Editor of IEEE Transactions on Communications from 2012 to 2017 and a Senior Editor of IEEE Communications Letters since January 2015 and IEEE Wireless Communications Letters from 2012 to 2016 and has been an Editor of IEEE Transactions on Wireless Communications since 2018. He has been an Area Editor of IEEE Open Journal of the Communications Society from 2019 to 2024.

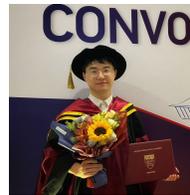

**Chao Ren** (Member, IEEE) received the B.E. degree from the School of Computer Science and Technology, Nanjing University of Aeronautics and Astronautics, Nanjing, China, in July 2017, and the Ph.D. degree from Interdisciplinary Graduate School, Nanyang Technological University, Singapore, in March 2022, holding Winner of Graduate College Research Excellence Award. Previously, he was a Wallenberg-NTU Presidential Postdoctoral Fellow at the School of Electrical and Electronic Engineering, Nanyang Technological University, Singapore. Currently, he is a Wallenberg-NTU Presidential Researcher at the School of Electrical Engineering and Computer Science, KTH Royal Institute of Technology, Sweden.

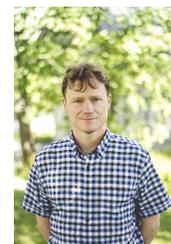

**Mikael Skoglund** (S'93-M'97-SM'04-F'19) received the Ph.D. degree in 1997 from Chalmers University of Technology, Sweden. In 1997, he joined the Royal Institute of Technology (KTH), Stockholm, Sweden, where he was appointed to the Chair in Communication Theory in 2003. At KTH he heads the Division of Information Science and Engineering. Dr. Skoglund has worked on problems in source-channel coding, coding and transmission for wireless communications, Shannon theory, information-theoretic security, information theory for statistics and learning, information and control, and signal processing. He has authored and co-authored more than 200 journal and 450 conference papers in these fields. Dr. Skoglund is a Fellow of the IEEE. During 2003–08 he was an associate editor for the IEEE Transactions on Communications. In the interval 2008–12 he was on the editorial board for the IEEE Transactions on Information Theory and starting in the Fall of 2021 he he joined it once again. He has served on numerous technical program committees for IEEE sponsored conferences, he was general co-chair for IEEE ITW 2019 and TPC co-chair for IEEE ISIT 2022. He is an elected member of the IEEE Information Theory Society Board of Governors.